# Connecting the timescales in picosecond remagnetization experiments

Marija Djordjevic, Markus Münzenberg*,

*IV. Physikalisches Institut, Universität Göttingen, Germany*

## Abstract

In femtosecond demagnetization experiments, one gains access to the elementary relaxation mechanisms of a magnetically ordered spin system on a time scale of 100 fs. Following these experiments, we report a combined micromagnetic and experimental study that connects the different regimes known from all-optical pump-probe experiments by employing a simple micromagnetic model. We identify spin-wave packets on the nanometer scale that contribute to the remagnetization process on the intermediate time scale between single-spin relaxation and collective precession.

A recent approach by B. Koopmans unifies the magnetization dynamics in ferromagnetic metals observed on the fs second timescale [1]. He connects the demagnetization time scale with the intrinsic energy-dissipation constant $\alpha$ by simple equations: the basic underlying idea is to extend the validity of the Landau-Lifshitz-Gilbert equation (LLG) that relates relaxation time and precession frequency into the fs timescale. $\alpha$ is inserted phenomenologically into the LLG equation, in analogy to a viscous Rayleigh energy dissipation function. Since $\alpha$ corresponds to an energy dissipation per precession period, the energy dissipation for a spin precessing at a frequency of a 100 THz in an exchange field in the order of $10^3$ T is large compared to the collective precession with a few MHz at a field of 10 mT. In the model he assumes for both time scales the same $\alpha$ and thus the same relevant relaxation channels. Generally, in metals the most relevant relaxation channel originates from Elliot-Yafet scattering. It connects the electron-spin coupling time constant inherently with the electron-lattice relaxation by introducing a certain spin-flip probability with each electron-lattice scattering event. Elliot and Yafet [2, 3] have investigated these spin-flip processes intensively in the 50ties on the theoretical side. They were followed later by experimental studies by Monod and Beuneu [4] for non ferromagnetic materials. The dominating spin-flip process in metals originates from the band mixing for symmetry points where two bands with indices $j$ and $j'$ come closely together in energy $\Delta E_{j,j'}$: due to the spin-orbit coupling $\lambda \mathbf{SL}$, the spin Eigen function does not interchange with the Hamiltonian any more which results in a the mixing of the spin-up and spin-down band. Consequently, a scattering event with a phonon, even without a spin wave involved, is now connected with a high spin-flip probability. In a ferromagnet, the result of these highly localized single spin-flip events are the so-called Stoner excitations. Fabian and Das Sarma reviewed the ideas in this field lately with realistic calculations for aluminum where these so-called spin-hot spots dominate the spin relaxation [5]. The spin-flip processes have also been identified in spin-diffusion experiments in the 80ties by Johnson and Silsbee [6], pioneering the new important research field spin currents and spin-torque switching [7]. To identify the Elliot-Yafet mechanism as a dominating elementary spin-flip mechanism in ultrafast demagnetization experiments, a high spin-flip probability for a scattering event originating from the spin-hot spots is needed [1, 8-11]. In the following, we will discuss an additional possible magnetic relaxation path within a micromagnetic toy model connecting the single-spin flip event with collective spin-wave excitations. The model will be verified by corresponding experiments.

In our ansatz, in the following we do not regard the details of the demagnetization process. The quenching of the magnetization due to spin-flip events is assumed instantaneous. As a result, a random magnetization is found within the penetration depth of the fs-laser pulse. As a first approximation, the exchange interaction is assumed non disturbed at this state. We can describe the exchange interaction, which originates from the details of the electronic structure, by one time independent variable $J_0$ averaging over details of the electronic structure and unit cell. The details of the spin scattering processes that cause an energy transfer from the spin system to the lattice as described in the introductory part, are represented by the averaged energy dissipation given by the damping constant $\alpha$. Both are included into the Landau-Lifshitz-Gilbert equation that governs the evolution for a micromagnetic model system implementing also temperature-excited fluctuations and spin disorder [12-14]. This concept is very successful to describe the static magnetization and spin dynamics in nanostructured films down to the ps range. In order to transfer the situation shortly after fs-laser pulse has reached the sample into an initial condition for our numerical approach, the demagnetization is transcribed by orienting the magnetization vectors locally by a random angle [15]. The degree of disorder generated by the perturbation is expected to be linear with the energy deposited in a certain depth. Thus the demagnetization profile mirrors the penetration profile of the laser light that decays exponentially on a length of 25 nm (termed as z-direction in the following). The degree of demagnetization corresponds to a given laser fluence in the experiment. Since the cell size determines the cutoff in frequency space, we made them as small as possible. As a drawback, the simulation had to be restricted to a cut through the demagnetization profile of the 50 nm thick Ni film in x-z-direction. The extension of the film in the y-direction is taken into account by an additional anisotropy in the film plane corresponding to the order of magnitude of the dipolar field of the Ni film ($K = -100 \times 10^3 \ J/m^2$, uniaxial in z-direction). To avoid artifacts due the restricted width we analyzed only a cross section of 50 nm x 100 nm from the total width of 200 nm. The damping constant $\alpha$ was chosen to be 0.08 corresponding to the experimental value [16].

For the micromagnetic simulation the OOMMF micromagnetic simulation code was used [17]. The initial situation is depicted in Fig. 1 (top). The pixel size of the color scale, representing the z-component of the magnetization, corresponds to the cell size of 0.5 nm. In the following, the degree of demagnetization is defined as the reduced magnetization at t= 0 averaged over the first 25 nm. For the simulation shown in Fig. 1 the degree of demagnetization was 55 %. The grainy profile and the gradual change of the demagnetization can be seen in the starting configuration (0 ps). As the time evolves, small domains are forming and spin waves are emitted from the excited surface. At a time scale of 700 fs, domains in the range of 3-6 nm have formed already. After 6.8 ps, the spin-wave front has penetrated the whole film thickness of 50 nm. The long wave-length excitations with a typical wave length of 15 to 20 nm dominate the upper part of the film, while in the lower part the dominating wave length is much smaller and in the order of 5 nm. This indicates a much higher spin-wave velocity for short wavelength excitations. In addition, the short wavelength excitations show a much shorter live time originating from their high precession frequency ω and therefore a high-energy dissipation. As the spin wave has traveled back and forth, standing waves are observed in Fig. 1 (16.7 ps).

Before discussing the evolution of the dominating spin-wave frequencies, decay and damping in more detail, a direct comparison with the experiment will show if the rather simple ansatz is successful to describe the basic phenomena observed in a fs-demagnetization experiment. As an experimental parameter the degree of pump fluence was increased systematically. The normalized experimental spectra for a polycrystalline Ni film Ni(50nm)/Cu(3nm)/Si(100) covered with 3 nm Cu capping layer are shown as the pump fluence is increased from 20 mJ/cm$^2$ to 60 mJ/cm$^2$, in Fig. 2 (left). The sample was characterized in detail [16]. The value of the Gilbert damping constant $\alpha$ used in the simulations corresponds to the value determined from the $k = 0$ precessional mode in the previous experiments. One observes for a higher degree of demagnetization a slower restoration time of the ferromagnetic signal. While for the lowest pump-fluence the relaxation occurs within a few ps and leads into a straight line, for very high fluences larger than 50 mJ/cm² the magnetic signal restores only slowly. On the right side of Fig. 2, the experiment is opposed to the micromagnetic simulation using the same procedure as in Fig. 1. The demagnetization, defined by the average demagnetization of the $M_x$ component over all cells over the first 25 nm, is varied from 1% to 75%. As the demagnetization is increased a slower restoring component is observed. The same tendency is observed for the experimental spectra: a slower restoring component adds significantly to the relaxation process if the demagnetization is increased to more than 55 %. The restoration can be divided into two time scales, analyzed with a double exponential model using two exponential decay times, $\tau_1$ and $\tau_2$, to characterize the relaxation processes. $\eta$ is the degree of demagnetization and $\gamma$ accounts for the size of the contribution of each component.

$$\frac{M}{M_0} = \eta \frac{e^{-\frac{t}{\tau_1}} + \gamma e^{-\frac{t}{\tau_2}}}{1+\gamma}, \quad t > 0 \qquad (1)$$

For the analysis of the experimental spectra, a constant background was added. This background takes into account the phonons contributing to the time-resolved Kerr rotation. A more detailed analysis of the spectra is described in [10]. The relaxation times extracted are given in Tab. 1. The ultrafast relaxation (time scale $\tau_1$) on a few 100 fs time scale has been investigated in many previous experiments. It varies from 420 to 780 fs, slightly increasing with the pump fluence. For pump fluences larger than 50 mJ/cm² the slower relaxation process (time scale $\tau_2$) ranges in between 4 to 7 ps and appears to have a major contribution within the same order of magnitude to the spectra.

| Fluence $[mJ/cm^2]$ | $\tau_1$ [fs] | $\tau_2$ [ps] |
|---|---|---|
| 20 | 420 | - |
| 30 | 450 | - |
| 50 | 800 | 4.4 |
| 60 | 780 | 7.1 |

Tab. 1. Analysis of the restoration of the time-resolved Kerr rotation $\theta_K$ in the experimental spectra (given in Fig. 2, left side) using two characteristic remagnetization time scales $\tau_1$ and $\tau_2$ as given in the text.

| Demagnetization [%] | $\tau_1$ [fs] | $\tau_2$ [ps] |
|---|---|---|
| 12 | 140 | - |
| 45 | 240 | - |
| 55 | 250 | 5.1 |
| 75 | 300 | 7.7 |

Tab. 2. Analysis of the restoration of the $M_x$ component from the micromagnetic model (spectra given in Fig. 2, right side) using two characteristic remagnetization time scales $\tau_1$ and $\tau_2$ as given in the text.

A comparison with Tab. 2 yields that the time scales observed in both cases (experimental and micromagnetic data) are almost identical. Solely $\tau_1$ is generally underestimated by the micromagnetic model. A reason might be that a reduced exchange interaction $J$ is expected shortly after excitation; experiments point to that [11]. This is not taken into account for within the micromagnetic model. Second, within the first 100 fs as long as the electron temperature has not equilibrated with the lattice, still further demagnetization due to spin flip processes can occur. Therefore, we expect an underestimation of the relaxation time $\tau_1$. Using a constant $J_0 \geq J(t)_2$ within the first 100 fs increases the energy dissipation within that time scale.

From the micromagnetic simulation, one gets an insight into the origin of the delayed remagnetization. Short wavelength spin waves and higher order spin-wave modes with a high spatial frequency dominate the dynamics within the first ps. These excitations are distributed very efficiently from the "hot"-spin region into the magnetically not disturbed region. The magnetization can be restored very fast by the emission of spin-wave packets and high-energy magnons [15]. High-energy magnons have been studied in detailed experimentally challenging neutron diffraction experiments in ferromagnetic metals in the 80ties, investigating the broadening of the spin-wave dispersion curves at high energies as predicted also at that time [18]. The broadening is related to a strong interaction with the Stoner band, called Landau damping: near the surface of the Brillouin zone, it results in a severe energy dissipation and extremely short live time of the high-energy spin waves. Recently, using spin-resolved electron energy loss spectroscopy [19] and theoretical investigation [20] these excitations have been related to spin-wave packets with lifetimes below 1 ps and localized within a few nanometers only. To discuss the remagnetization process and nature of the excitations in further detail, vertical cuts through the film showing the magnetization profile for $M_x$, $M_y$ and $M_z$ components (averaged over four cells) are shown in Fig. 3. They give a detailed insight into the remagnetization process. Two competing relaxation scenarios are possible and will be distinguished in the following by analyzing the spatial frequencies. In the first scenario, the higher frequency of the short spin-wave packets alone explains the higher damping. As stated in the introductory part, $\alpha$ is related to the energy dissipation per period. The magnetic spin-wave modes are not interacting within this model. In the second scenario, energy is transferred from the highest excited mode into modes owning a lower energy following the spin-wave dispersion towards the $k=0$ precessional mode [16, 21-23]. On the right side in Fig. 3, the corresponding Fourier transform of the vertical lateral cuts through the magnetization profiles is shown. In our micromagnetic model, spin-wave excitations larger than approximately four times the simulation cell size are accessible (spatial frequency $k_z/2\pi = 0.5 \ nm^{-1}$). The spectra are shifted vertically for clarity. High spatial frequencies dominate after excitation at t=0. During the relaxation process, the center of the spectral weight moves towards lower spatial frequencies. For 6.7 ps the center is at around $k_z/2\pi = 0.1 \ nm^{-1}$. This corresponds to a spatial spin-wave period of 10 nm.

Both scenarios predict the evolution of the frequency spectrum as a function of time. The overall damping results in an overall reduction of the spectral amplitudes. It is especially efficient for the high spatial frequencies that are reduced in amplitude first. But also, a shift of the spectral weight to lower frequencies is observed. This is an indication for an energy transfer from high-energy modes to low-energy modes within the spin-wave relaxation chain [24]. Herewith it is possible to speculate about the starting point of the relaxation chain: an energy transfer from high-energy excitations to the lower-order spin-wave excitations is expected since the high-energy excitations are overpopulated in an ultrafast demagnetization experiment. Extrapolated to the basic excitation standing at the beginning of the relaxation chain, it seems plausible that the Stoner excitations, that are populated by the elementary relaxation processes of the hot electron system as described in the introductory part, decay into short wave length spin excitations and gradually relax to the lower spatial frequency excitations.

In summary, we introduce a micromagnetic model that describes the general features and time scales observed in an ultrafast demagnetization experiment. By increasing the demagnetization, two relaxation time scales $\tau_1$ and $\tau_2$ can be identified. The additional time scale $\tau_2$ is an experimental evidence for short wavelength spin-wave excitation owning a live time of a few ps. The micromagnetic model gives an insight into the details of the magnetic relaxation channels. We propose a relaxation model, where energy is transferred from high-energy magnetic excitations to low energy spin wave exitations within a spin-wave relaxation chain. Extended future models may also include a variation of $J(t)$. Adjusting $J(t)$ in a way that the resulting magnetization $M(t)$ describes the experimental data will vice versa give an insight into the temporal evolution and quenching of the exchange interaction within the first 100 fs.

***

Support by the Deutsche Forschungsgemeinschaft within the priority program SPP 1133 is gratefully acknowledged.

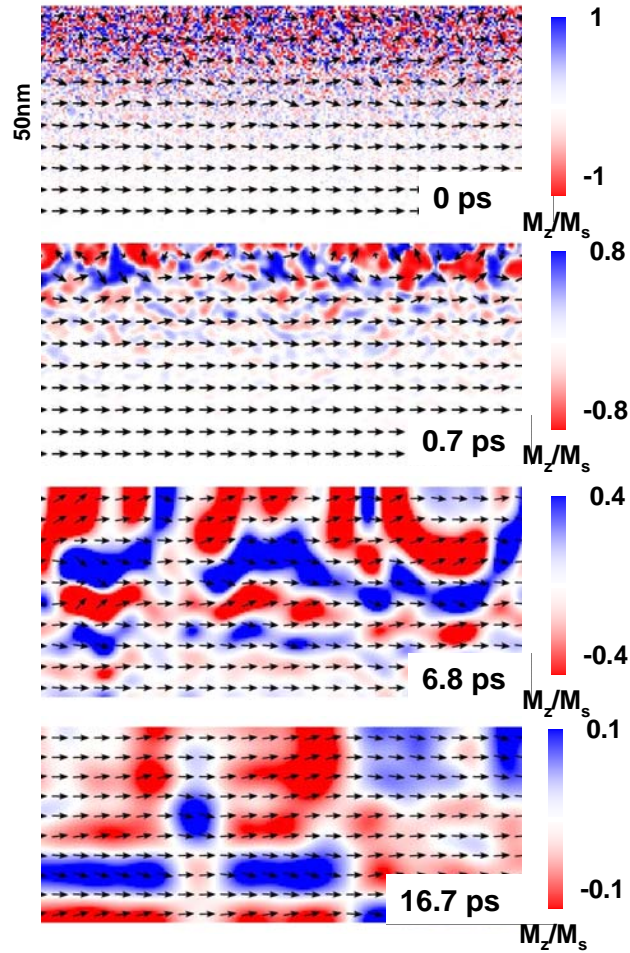

Fig. 1. Micromagnetic simulation for a Ni film with t=50 nm and 55 % demagnetization. The evolution of the remagnetization is shown after demagnetization by a fs-laser pulse. The demagnetization profile mirrors the penetration profile of the laser light that decays exponentially on a length of 25 nm.

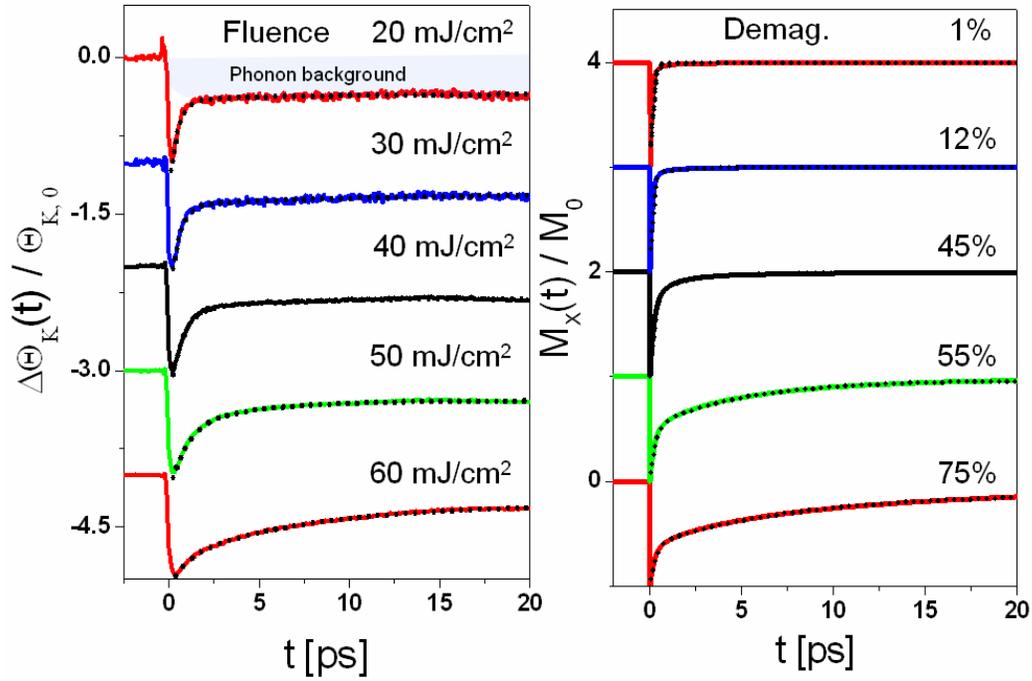

Fig. 2. Left side: time-resolved Kerr rotation for a 50 nm thick Ni film after demagnetization with increasing pump-pulse fluence. Right side: micromagnetic simulation of the relaxation of the magnetization for a 50 nm thick Ni film with increasing demagnetization. The dotted line is an analysis using a double exponential function. The spectra are normalized and vertically shifted for comparability.

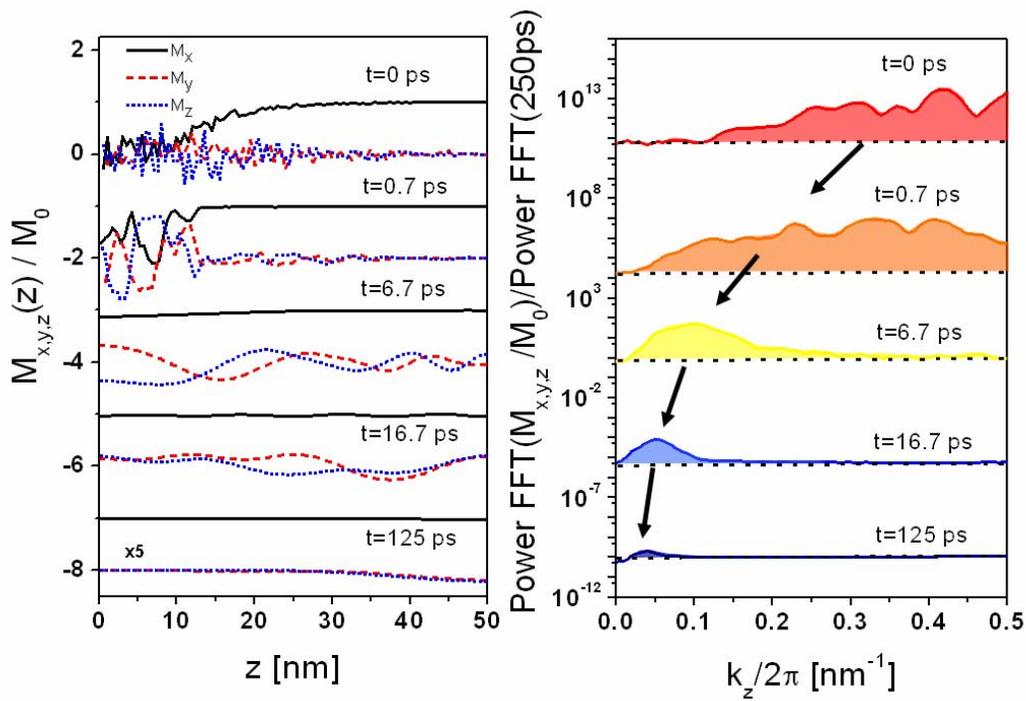

Fig. 3. Cut through the micromagnetic simulation for the Ni film with t =50 nm and 55 % demagnetization. On the left, the evolution spin wave emission from the excited area is shown in real space. On the right, the corresponding Fourier transform is shown as a function of the spatial frequency.